\documentclass[12pt,a4paper]{article}
\usepackage[utf8]{inputenc}
\usepackage[french,english]{babel}
\usepackage{authblk}
\usepackage{multicol}
\usepackage{hyperref}
\hypersetup{
  colorlinks = true,
  linkcolor = red,
  citecolor  = red,
  urlcolor = blue
}

\usepackage{graphicx}
\usepackage{listings}
\usepackage{amsmath}
\usepackage{array}

\newcommand{\ta}[1]{(table~\ref{#1})}
\newcommand{\fig}[1]{figure~\ref{#1}}
\newcommand{\code}[1]{code~\ref{#1}}

\title{Caractérisation de maliciels Android basée sur l'isomorphisme de sous-graphes}
\author[1]{Alain Menelet}
\author[2]{Charles-Edmond Bichot}
\affil[1]{Armée de l'air, ESIOC, France}
\affil[2]{Armée de l’air, CNRS, LIRIS, École de l'air, CREA, France}

\date{Juin 2020}

\begin{document}

\maketitle

\selectlanguage{french}
\begin{abstract}
  Le succès grandissant de l'écosystème Android s'est accompagné d'un réel engouement pour toute une génération de développeurs peu scrupuleux qui y voit là un terrain de jeu lucratif. L'année 2010 marque l'explosion des maliciels sur Android et ces 7 dernières années ont été le témoin d'une ingéniosité de plus en plus importante afin de rendre ces programmes de plus en plus indétectables. Nous présentons dans cet article une nouvelle méthode permettant de caractériser de manière très précise si un logiciel Android fait parti d'une famille de maliciels. Cette méthode consiste dans un premier temps à obtenir une signature très précise des maliciels en se basant sur leur bytecode. Puis, par recherche d'isomorphisme de sous-graphe, elle permet d'identifier si un programme Android fait parti d'une famille de maliciels dont on possède la signature. 
Nous avons testé notre méthode sur deux jeux de données, un jeu de données de laboratoire construit de manière ad hoc et un jeu de données réelles. Les résultats obtenus valident notre approche.
\end{abstract}

\selectlanguage{english}
\begin{abstract}
The Android operating system is the most spread mobile platform in the world. Therefor attackers are producing an incredible number of malware applications for Android. Our aim is to detect Android's malware in order to protect the user. To do so really good results are obtained by dynamic analysis of software, but it requires complex environments. In order to achieve the same level of precision we analyze the machine code and investigate the frequencies of ngrams of opcodes in order to detect singular code blocks. This allow us to construct a database of infected code blocks. Then, because attacker may modify and organized differently the infected injected code in their new malware, we perform not only a semantic comparison of the tested software with the database of infected code blocks but also a structured comparison. To do such comparison we compute subgraph isomorphism. It allows us to characterize precisely if the tested software is a malware and if so in witch family it belongs. Our method is tested both on a laboratory database and a set of real data. It achieves an almost perfect detection rate.
\end{abstract}

\section{État de l'art}

Android occupe aujourd'hui près de 86\% des parts du marché des ordiphones (\emph{smartphone}) avec la version 6.0 présente sur plus de 30\% des équipements actuels. Plus de 99\% des maliciels (\emph{malwares}) conçus pour attaquer les dispositifs mobiles ciblent les appareils Android~\cite{fsecuresecurity2017}. Le nombre de maliciels et leur ingéniosité ne cessent de croître. Ils combinent des mécanismes tels que l'obfuscation, le chiffrement ou bien la communication avec leur centre de commandes et contrôle afin d'échapper au système de détection. Le Google Play Store doit lui aussi faire face à la recherche de menace. Lorsqu’une application est créée à destination du store de Google, elle doit être signée avec la clé privée du développeur puis examinée par un système de détection appelé Bouncer. Mis en place en 2011, il répond à la croissance importante des maliciels disponibles sur le store officiel. Peu d’informations sont disponibles sur son mode de fonctionnement mais nous savons qu’il procède au scan des applications récentes et anciennes et selon Google c’est 40\% d’applications malveillantes qui ont ainsi été supprimées peu de temps après sa mise en place. Deux chercheurs en sécurité ~\cite{youtuBouncer} ont affirmé pouvoir l’identifier et le contourner. Pour cela, ils ont modifié une application en ajoutant un shellcode afin de contourner cette vérification et ainsi en apprendre un peu plus sur ce système via un shell distant.

Début 2017, Google annonce Verify Apps, un système de protection qui vérifie le comportement des applications. Si l’application est considérée comme une application potentiellement dangereuse, Verify Apps informe l’utilisateur afin de procéder à sa suppression. Selon Google, Verify Apps analyserait plus de 750 millions par jour. Si aucune information précise n'est donnée par Google concernant cette vérification, il semblerait qu'il s'agisse d'une analyse aussi bien statique que dynamique. Il a été montré que les applications vérifiées sur Google Play étaient exécutées dans un machine virtuelle dont il est possible de contruire le profil~\cite{Oberheide}. Cependant, ce système de protection n'est pas infaillible et comme le montre l'exemple du maliciel Dvmap~\cite{dvmap} qui après obtention des droits root désactivait Verify Apps et passait ainsi inaperçu.

La détection de maliciel se fait habituellement en utilisant des signatures, en se basant sur les \emph{apk}, ou des profils comportementaux. Signatures et profils peuvent être construits en se basant sur une analyse statique ou dynamique du maliciel. L'analyse statique permet de construire des signatures de maliciels pour ensuite comparer la signature d'un nouvel échantillon à celles-ci~\cite{griffin2009automatic,feng2014apposcopy}. 

Dans la méthode que nous proposons, nous cherchons à décrire le profil comportemental d'une application Android. Souvent, l'étude du profil comportement est fait de manière dynamique. Pourtant, nous considérons que l'analyse statique permet aussi une bonne représentation de ce profil, en considérant tous les embranchements possibles, contrairement à l'approche dynamique qui pourrait être contournée par le maliciel~\cite{Abraham2015}. En effet, ceux-ci mettent en oeuvre des techniques de contournement comme la transformation pour échapper aux signatures~\cite{Rastogi2014} ou la détection de bac à sable~\cite{Vidas:2014:EAR}. Ainsi, nous proposons une approche statique se basant sur les blocs de bytecodes constituants l'application. Un graphe de flots de contrôle est construit à partir des enchaînements possibles entre ces blocs.

D'autres méthodes existent, se basant sur l'API, pour détecter les maliciels sous Android. Dans ~\cite{mamadroid}, Mariconti \emph{et all} étudient les séquences d'appels des API afin de determiner le comportement de l'application. Ils ont pour cela procédé à l'extraction des graphes d'appels et ont utilisé la chaîne de Markov afin de représenter les séquences d'appels. Ils ont obtenu un taux de détection intéressant (99\%) sur des nouveaux maliciels issus de la même période que les maliciels ayant servi à l'apprentissage. Cette valeur décrois au fur et à mesure que nous nous éloignons de cette période d'apprentissage. Dans~\cite{li2015detection}, Li \emph{et all} procédent à l'extraction des API afin de déterminer les dépendances caractéristiques. S'appuyant sur des algorithmes d'apprentissage automatique notamment via des Forêt d'arbres décisionnels ils ont pu identifier près de 96\% des maliciels présents dans leur base de données. Ces travaux montrent qu'il y a une étroite relation entre les API et le comportement du maliciel. Cette relation est inhérente à l'architecture d'Android. Dans~\cite{gascon2013structural}, Gascon \emph{et all} présentent une étude sur la pertinence des graphes d'appels de fonctions dans la recherche de variantes. Par apprentissage automatique via machine à vecteur de support sur plus de 12\,000 maliciels, ils ont obtenu 89\% de détection contre 1\% de faux positifs. Très récemment

\section{Construction de notre approche}

\subsection{Intérêt de la méthode proposée}

Dans cet article, nous présentons une nouvelle approche de détection de maliciels par isomorphisme de sous-graphe en travaillant sur les blocs d'opcodes. Utiliser le problème de l'isomorphisme de sous-graphe nous permet de réaliser une comparaison structurée entre les blocs d'opcodes en plus d'une comparaison sémantique. C'est le premier intérêt de notre approche comparativement à l'état de l'art qui ne procède généralement qu'à une comparaison sémantique. 

L'isomorphisme de sous-graphe va nous permettre d'évaluer la ressemblance d'un logiciel à analyser avec un maliciel connu. Plus précisément, l'algorithme de calcul d'isomorphisme de sous-graphe va comparer les graphes des blocs d'opcodes d'un logiciel à analyser à une base de donnée de graphes de blocs d'opcodes de maliciels.
C'est le second intérêt de notre méthode comparativement à l'état de l'art qui ne descend pas le plus souvent à un tel niveau de détail : le langage machine.

Comme nous venons de le voir, nous travaillons sur le langage machine et plus précisément les codes opération (opcode). Nous considérons que, comme pour tout langage, le vocabulaire utilisé est révélateur des intentions de celui qui l'utilise. 
Nous allons donc considérer que les blocs d'opcodes issus d'un code chargé possèdent un vocabulaire légèrement différent des autres blocs d'opcodes.
Dans notre approche, nous utilisons la fréquence relative des opcodes en suivant une méthode 
de fouille de textes (text mining). C'est le troisième intérêt de notre méthode, elle analyse le langage machine pour en extraire des connaissances de similarité avec du code chargé.

Nous partons de l'hypothèse qu'un développeur de maliciel suit les bonnes pratiques du génie logiciel et va donc réutiliser du code existant pour développer un maliciel. C'est le principe de la signature qui permet de retrouver des blocs de code machine identiques. Cette hypothèse est particulièrement vrai pour les maliciels qui sont souvent des compositions de programmes sains et de morceaux de code chargés. Cependant, afin d'être plus précis dans la détection, il faut pouvoir tenir compte du fait qu'un développeur de maliciel va adaptés, modifiés, complétés, en un mot transformer les morceaux de code chargé qu'il utilise.
Nous proposons une approche qui permet de lever ce verrou. C'est le quatrième intérêt de notre méthode : elle est robuste aux transformations du code chargé. Notre approche permet donc de retrouver plusieurs maliciels d'une même famille.

\subsection{Présentation globale de notre méthode}

\begin{figure}
  \includegraphics[scale=0.3]{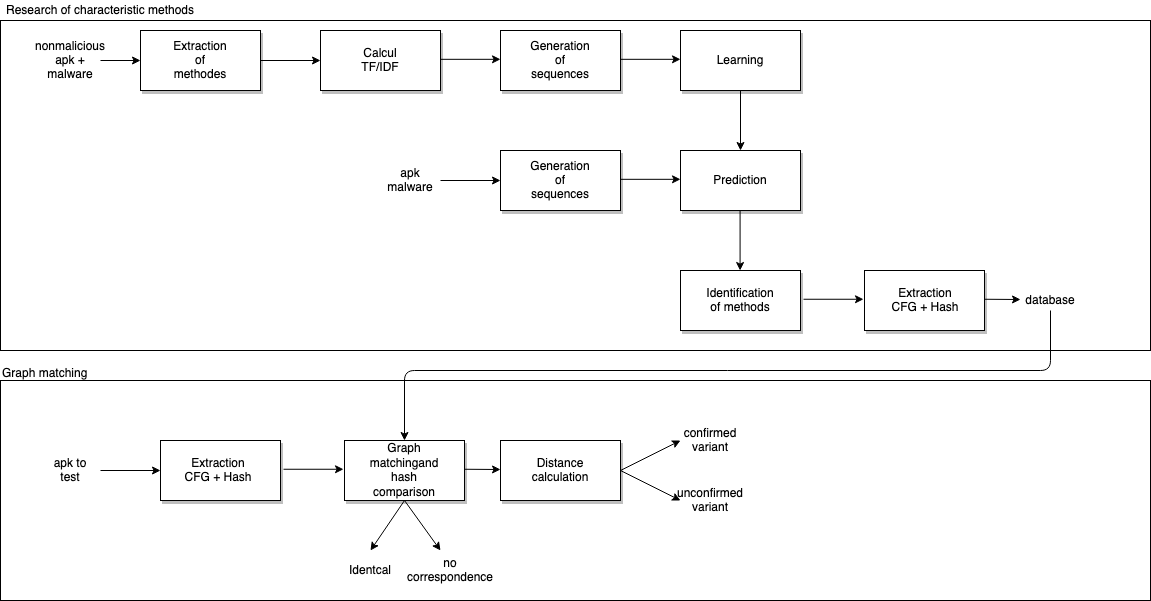} 
  \caption{Présentation global de notre approche}
  \label{global}
\end{figure}

La figure~\fig{global} présente les différentes étapes de notre approche.

Afin de détecter si un logiciel est sain, notre méthode repose sur la comparaison entre ce logiciel et une base de donnée de maliciels. Plus précisément des graphes de flots de contrôle (CFG) sont créés à partir du code machine du logiciel à analyser. La structure du logiciel et de ses enchaînements logiques est ainsi préservée. Les noeuds du graphe sont étiquetés avec le hash du bloc d'opcode sous-jacent. Une description plus détaillée de la construction des CFG est présenté en section~\ref{CFG}.

La comparaison entre les CFG du logiciel à analyser et ceux de la base de donnée est réalisée en calculant leurs isomorphismes de sous-graphe. Nous présentons les détails de ce calcul en section~\ref{isomorphisme}. 
Le résultat du calcul nous indique pour chaque pair de graphe comparé : soit que ces deux CFG sont identiques et dans ce cas nous savons que le logiciel à analyser est un maliciel connu ; soit qu'il n'y a aucune correspondance entre les deux CFG ; soit qu'il y a une correspondance et dans ce cas on calcul la distance entre ces deux CFG afin de savoir si l'on est en présence d'une variante d'une famille de maliciel connu. 

La base de donnée de CFG est un élément critique de notre méthode.
Cette base de donnée est constituée en partant d'une base de donnée de maliciels connus. 
Pour chacun de ces maliciels, nous allons en extraire ses CFG les plus caractéristiques. Bien évidement, ceux-ci peuvent ne pas correspondre à des séquences de code chargé, mais ils doivent permettre de caractériser le mieux possible ce maliciel comparativement aux autres CFG de celui-ci.
Afin de la constituer cette base de donnée de CFG, nous utilisons l'apprentissage automatique.
Cet apprentissage va permettre d'identifier les séquences de code machine les plus caractéristiques des maliciels. Une fois les séquences de code machine caractéristiques identifiés, elles seront transformées en CFG pour alimenter la base de donnée de CFG.

Comme toute approche utilisant l'apprentissage automatique, il faut dans un premier temps apprendre un modèle à partir d'une base de donnée d'apprentissage pertinente puis dans un second temps réaliser la détection. Nous utilisons la fouille de textes pour notre apprentissage.
Nous allons nous intéresser à la fréquence relative des opcodes afin de discriminer les blocs de code chargés.
Ainsi l'apprentissage se fait à partir de vecteurs caractéristiques issus de l'analyse de la fréquence TF/IDF des opcodes dans les logiciels. Cette analyse est présentée en section~\ref{bdopcodes}. 

Une fois le modèle appris à partir d'une base de donnée de logiciels sains et de maliciels pour lesquels la fréquence TF/IDF des opcode a été extraite, nous pouvons prédire dans les maliciels que nous connaissons quels sont les séquences de code chargées.

\section{Graphes de flot de contrôle}
\label{CFG}

Nous présentons dans cette section la création des graphes de flot de contrôle, \emph{Control Flow Graphs} (CFG) en anglais. La création de ces graphes constitue la première étape de notre démarche.

Notre approche consiste à décrire le profil comportemental d'une application Android. Ce profil comportemental est caractérisé de manière statique par les blocs de bytecodes constituants l'application et leurs possibles embranchements. Un graphe de flot de contrôle est ainsi créé dont les sommets sont les empreintes numériques, les hashs, des blocs de bytescodes.

Un flot de contrôle représente l'ordre dans lequel les instructions d'un programme sont exécutées. Cette modélisation a été introduite par Frances Allen \cite{Allen:1970} qui représente le flot de contrôle comme un graphe orienté dont les noeuds sont des blocs élémentaires et les arcs les enchainements possibles entre ces blocs. Ces graphes orientés, ces CFG, peuvent servir aussi bien à une analyse dynamique que statique d'un programme.

Le plus souvent, les CFG sont construits à partir des appels aux API. L'originalité de notre approche est double puisque nous nous basons sur les mnémoniques du bytecode et que nous marquons les sommets du CFG par l'empreinte numérique des blocs de bases.

\subsection{Création des blocs de base}

Un bloc de base est composé de plusieurs mnémoniques, \emph{i.e.} instructions. Pour chaque instruction, nous allons déterminer s'il s'agit de la première instruction, d'un embranchement conditionnel ou bien de l'instruction de fin de fonction.

Nous nous sommes appuyés sur la bibliothèque smali développé par Ben Gruver~\cite{smali} afin de procéder à l'extraction des opcodes. Cette bibliothèque est un assembleur/désassembleur pour le format dex d'Android.

Prenons par exemple la fonction z.run() extraite du maliciel de la famille DroidJack qui est composée des 15 instructions suivantes.

\begin{lstlisting}[caption=Dalvik bytecode de la fonction z.run(), label=a_label, captionpos=b]
IGET_OBJECT
CONST_4
AGET_OBJECT
IGET_OBJECT
CONST_4
AGET_OBJECT
CONST_4
IF_EQZ
CONST_STRING
INVOKE_VIRTUAL
MOVE_RESULT_OBJECT
INVOKE_STATIC
MOVE_RESULT_OBJECT
SPUT_OBJECT
RETURN_VOID
\end{lstlisting}

Chaque instruction conditionnelle va déterminer la fin d'un bloc. Ce type d'instruction va déterminer deux possibilités. Soit la condition est respectée soit elle ne l'est pas. Il y a donc au maximum deux possibilités d'embranchement. Par exemple, prenons la ligne ci-dessous

\begin{lstlisting}[caption= structure d'une instruction bytecode conditionnel, captionpos=b,language={[x86masm]Assembler}]
3902 1200 if-nez v2, 0014
\end{lstlisting}

3902 1200 correspond aux opcodes\footnote{Les opcodes sont les mnémoniques, \emph{i.e.} les instructions, de base du bytecode Dalvik} et if-nez au Mnemonique. Si la valeur stockée dans v2(8 bits) est différente de 0 alors nous sautons à l'adresse actuelle (offset) + 0x12 sinon le branchement se fait à l'adresse 0014.

Les instructions ci-dessus vont donc être décomposées en 3 blocs de base. Le nombre de destinations pour chaque bloc dépend du type d'opcode; dans le cas d'un if nous obtenons deux sorties alors que pour le mnemonique packed-switch payload qui est l'équivalent d'un switch en langage C, il saute vers une nouvelle instruction en fonction de la valeur d'un registre qu'il compare. Les valeurs de sauts sont stockées dans une table d'offsets.
Nous procédons ensuite au découpage des 15 instructions en blocs de base.

La première adresse est égale à 0 et est considérée comme point d'entrée. Les octets sont lus puis traduits en mnemonique. 

L'instruction invoke\_virtual (0x6E) n'est pas la première instruction, ne fait pas partie de la liste des instructions permettant des embranchements, possède un poids de 3 ce qui veut dire que la prochaine instruction sera à l'adresse 54. Ces éléments définis, nous pouvons découper notre fonction en 3 blocs, des adresses 0 à 11, puis de 13 à 18 et de 19 à 25.

\begin{figure}[h]
  \centering
  \includegraphics[scale=0.3]{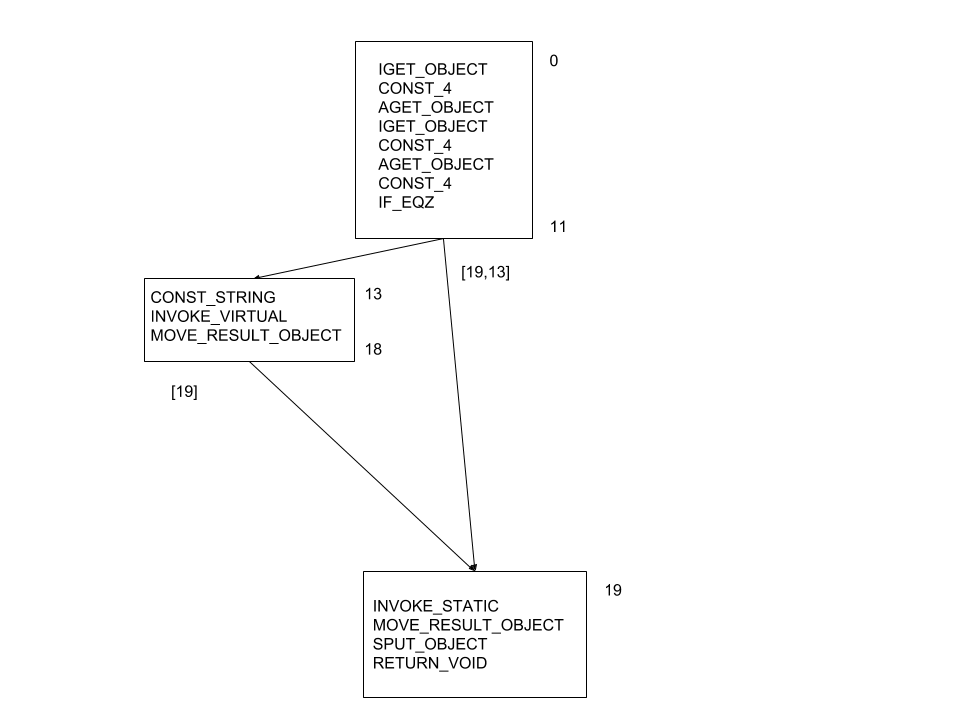} 
  \caption{\label{zrun}\hspace{0.15cm}Graphe de flot de contrôle de la fonction z.run()}
\end{figure}

Tous ces éléments nous permettent de créer le graphe de flot de contrôle (CFG) de la fonction z.run() représenté~\fig{zrun}. Ce CFG n'est pas encore marqué. Nous allons voir comment le marquer ci-dessous.

\subsection{Empreintes numériques des blocs de base}


Prenons comme exemple un bloc composé de trois mnémoniques : \texttt{NEW\_INSTANCE}, \texttt{INVOKE\_DIRECT} et \texttt{SPUT\_OBJECT}. Pour obtenir l'empreinte numérique, \emph{i.e.} le hash du bloc, nous concaténons les caractères de ces trois mnémoniques et calculons le hash MD5 de cette chaîne. Le choix de MD5 est une simple question de taille pour alléger les différents traitements par rapport à du SHA256. La probabilité d'une collision MD5 n'est pas considérée comme suffisamment importante pour privilégier un autre système de hash. Ainsi, le hash du bloc précédent vaut \texttt{d459c7ade0c9fe2c73e7c73f08c090a}. 

Cette méthode permet de connaître les opcodes ainsi que l'ordre dans lequel ils sont rangés. La bibliothèque smali 
permet également d'obtenir une granularité plus importante telle que la valeur des registres et les différentes adresses utilisés. Cela présente néanmoins l'inconvénient de prendre en considération l'emplacement mémoire des variables ce qui aurait pour conséquence de modifier les hashs et fausserait les résultats.


\subsection{Création du CFG marqué}

Nous allons voir dans cette section comment nous obtenons le CFG final dont nous nous servirons pour l'étape de classification/caractérisation.


Bien que le CFG soit représentatif d'un comportement, les précédents travaux présentent l'inconvénient de renvoyer beaucoup trop de faux positifs. En effet il est très commun d'avoir le même CFG pour deux fonctions différentes tout simplement car l'agencement des blocs de base est identique. Nous avons pallié cette situation en marquant avec un identifiant unique chaque bloc de base afin de rendre cet enchaînement unique. La marque en question est l'empreinte numérique du bloc de base tel que nous l'avons présenté ci-dessus.

La~\fig{zrun} représente le CFG de la méthode z.run(). Il suffit maintenant de calculer l'empreinte numérique de chacun des trois blocs de base. 
Nous obtenons à la fin de ce processus une signature unique de la fonction z.run().
Cette signature est constituée d'une part de la matrice d'adjacence du CFG et d'autre part des empreintes numériques de ses noeuds. Il s'agit de la signature du CFG marqué. 
Le~\code{signaturez} présente les empreintes numériques pour la méthode z.run().

\begin{lstlisting}[caption= Signature de la m\'ethode z.run(),captionpos=b,label=signaturez]
Lnet/droidjack/server/z.run()V;3;0:1,2;1:2;
[dc50d4bb08e48f0906a88db23709e414, 
f53a5540b26a20a907b99d934d2e704a,
fe467c14a6caa33801990f9c094b52ae]
\end{lstlisting}

\section{Isomorphismes de sous-graphes}
\label{isomorphisme}

Le CFG créé précédemment est  caractéristique du comportement propre à un programme ou à un programmeur. Ces derniers étant bien souvent peu enclins à réécrire à chaque fois un maliciel, ils préfèrent se baser sur l'existant. Certaines fonctions caractéristiques pourront être les témoins de la présence d'un maliciel ou d'une de ses variantes. 
Au regard des différents cas présentés ci-dessus, la construction d'un CFG marqué basé sur des blocs de bases de mnémoniques de bytecodes permet de répondre de manière satisfaisante à notre approche, principalement parce que la table des numéros de ligne, issues comme nous l'avons vu de la compilation, n'est pas disponible dans le cas d'un apk téléchargé d'Internet.

Afin de classifier un programme inconnu comme étant sain ou infesté, il faut comparer la signature de celui-ci, c'est-à-dire de ses CFG marqués, avec un dictionnaire de signatures de maliciels. Cette comparaison se fait par la recherche d'isomorphismes de sous-graphes entre la signature du programme inconnu et une signature du dictionnaire. Nous ne procédons à ce calcul d'isomorphisme que dans le cas où au moins une empreinte numérique correspond entre CFG du programme candidat et du dictionnaire. 

\subsection{Recherche d'isomorphismes de sous-graphes}

Nous allons nous appuyer sur la théorie des graphes en cherchant à déterminer s'il existe un couplage possible entre les deux CFG marqués. 
À partir de la signature du programme inconnu et d'une des signatures du dictionnaire, nous créons une matrice de correspondance qui servira de base au calcul du couplage. 
Par la suite, nous désignerons la signature du programme inconnu comme étant la signature candidate et la signature issu du dictionnaire comme étant la signature réfèrent.

\begin{figure}
  \centering
  \includegraphics[width=\linewidth]{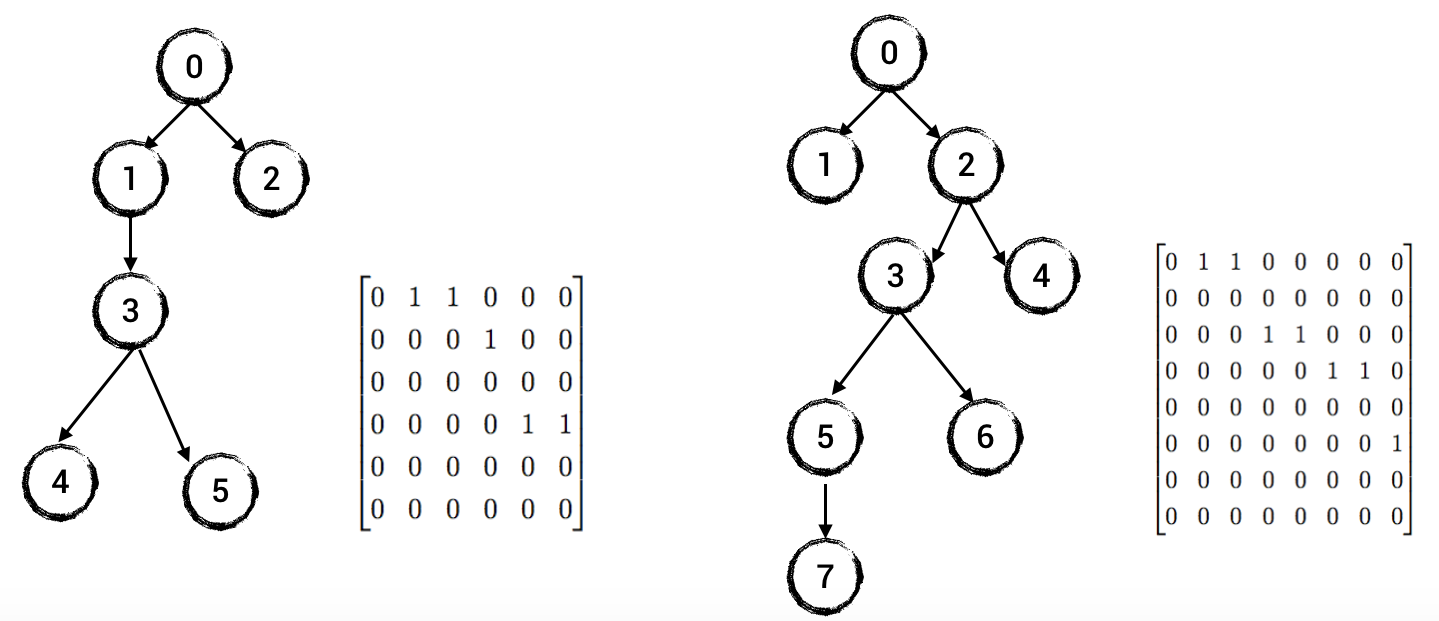} 
  \caption{Comparaison des CFG référent (gauche) et candidat (droite)}
  \label{g1g2}
\end{figure}

La recherche d'isomorphisme de sous-graphe entre les deux CFG candidat et référent est calculé grâce à l'heuristique suivante. 
Pour chaque noeud du CFG réfèrent il faut déterminer quels sont les noeuds du CFG candidat qui respectent les critères de viabilité. Ces critères sont que les noeuds doivent être à la même profondeur par rapport au sommet, le nombre d'arcs entrant doit correspondre, les arcs sortant sont au moins égaux à ceux de la signature et enfin le noeud père doit également être candidat. Le choix de l'ordre des noeuds à analyser se fait en fonction du \textbf{nombre d'enfant que chacun d'entre eux possède en commencant par celui qui en possède le plus}. L'heuristique commence par les noeuds d'entrée des CFG. 
La~\fig{g1g2}~présente le CFG référent et le CFG candidat ainsi que leurs matrices d'adjacences.

\begin{figure}
  \centering
  \includegraphics[scale=0.4]{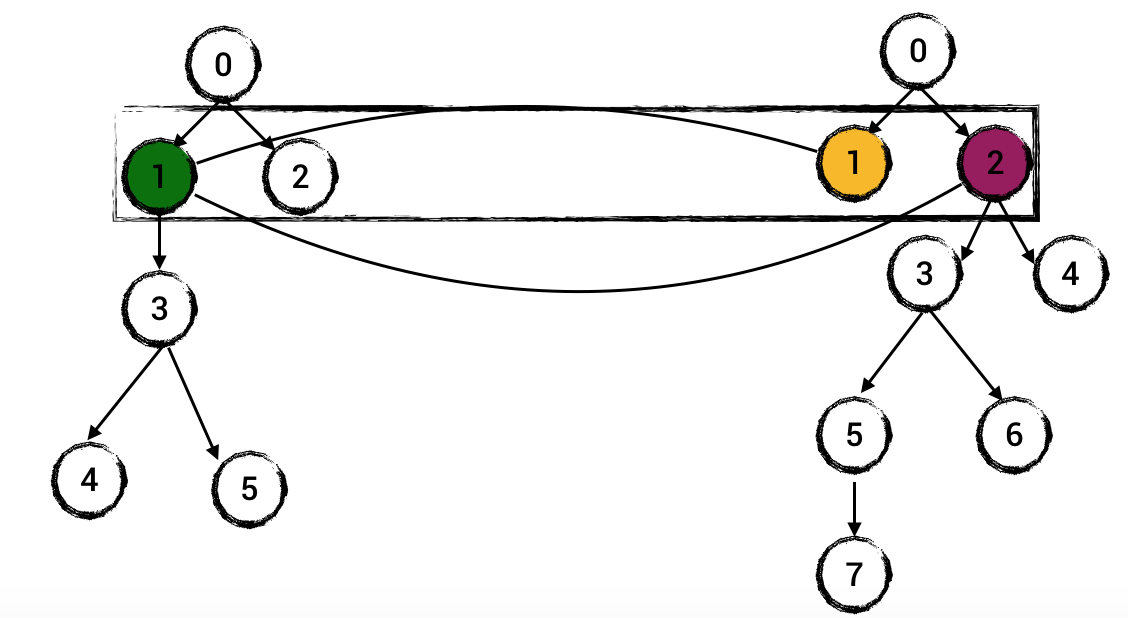} 
  \caption {Correspondances possibles entre le noeud 1 du CFG référent et les noeuds du CFG candidat}
  \label{candidat}
\end{figure}

\begin{figure}
  \centering
  \includegraphics[scale=0.2]{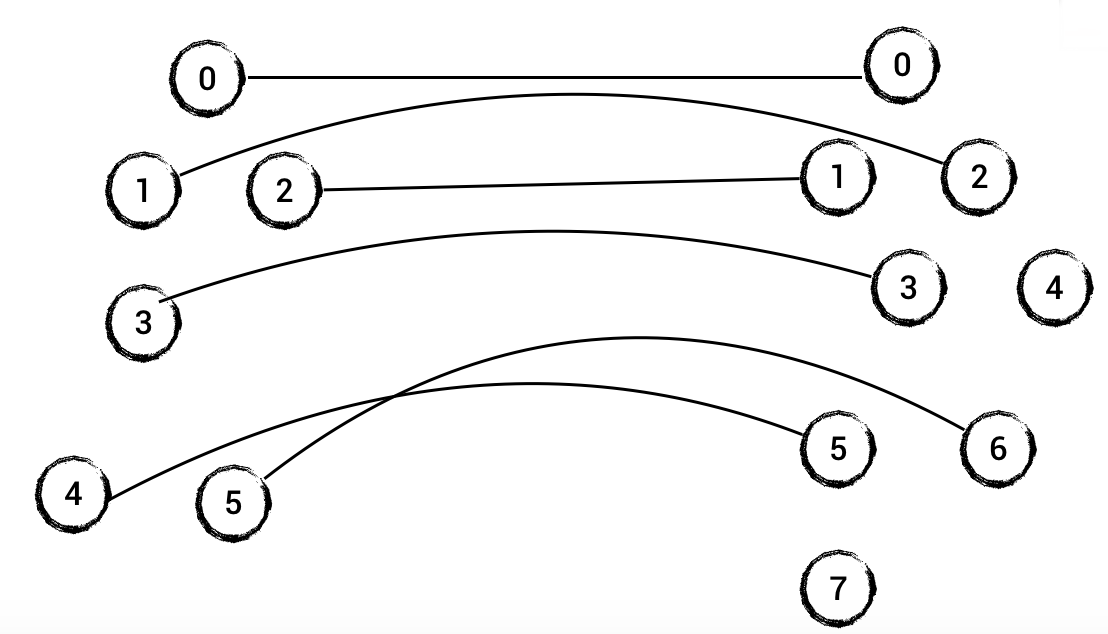} 
  \caption {Correspondance finale entre le CFG référent et le CFG candidat}
  \label{biparti}
\end{figure}

La matrice de correspondance est de forme $m\times{}n$, $m$ la taille du CFG référent et $n$ la taille du CFG candidat. La matrice de correspondance représente la concordance entre les deux CFG. Dans le cas de~\fig{g1g2}, le référent a une matrice de $6\times{}6$ et le candidat de $8\times{}8$, la matrice de correspondance sera donc de $6\times{}8$. L'heuristique commence par le noeud d'entrée soit $0$, au même niveau dans les deux graphes, il s'agit du point d'entrée. Il n'y a pas d'arc entrant et il possède deux sorties. La matrice de correspondance va donc avoir en position $(0,0)$ la valeur 1. Le noeud 1 du CFG référent peut être associé au noeud 2 du CFG candidat car il a le même nombre d'arc entrants et le noeud père est aussi un noeud candidat,~\fig{candidat}. Il est donc ajouté à la matrice de correspondance. Cette heuristique génère également potentiellement des doublons qu'il est nécessaire d'éliminer. Dans le cas du noeud 2 du CFG référent, les noeuds 1 et 2 du CFG candidats sont des choix potentiels. Cependant comme le noeud 2 du CFG candidat est déjà associé au noeud 1 du CFG référent, il est déjà pris. Une fois l'heuristique appliquée aux deux CFG, nous obtenons les correspondances entres sommets de la~\fig{biparti}. La matrice de correspondance de notre exemple vaut alors :
\[
  \begin{bmatrix}
    1 & 0 & 0 & 0 & 0  & 0 & 0 & 0\\ 
    0 & 0 & 1 & 0 & 0 & 0 & 0 & 0\\ 
    0 & 1 & 0 & 0 & 0 & 0 & 0 & 0\\ 
    0 & 0 & 0 & 1 & 0 & 0 & 0 & 0\\ 
    0 & 0 & 0 & 0 & 0 & 1 & 0 & 0\\ 
    0 & 0 & 0 & 0 & 0 & 0 & 1 & 0
  \end{bmatrix}
\]

En dernière étape, nous appliquons l'algorithme d'Hopcroft–Karp qui permet dans le cas d'un graphe biparti de trouver un couplage de cardinal maximum en temps polynomial. Il s'agit de connaître le nombre d'arcs qui ne partagent pas un sommet et si ce nombre est égale au nombre de sommet du graphe signature. Nous obtenons alors le résultat du calcul d'isomorphisme de sous-graphe recherché. Ce résultat est affiné par la correspondance entre empreintes numériques

\subsection{Correspondance entre empreintes numériques}

Une fois qu'un isomorphisme de sous-graphe a été trouvé, nous vérifions la correspondance entre les empreintes numériques de leurs sommets. Pour cela, nous prenons le CFG référent et pour chaque noeud de celui-ci, nous regardons le noeud correspondant du CFG candidat et vérifions la correspondance entre leurs empreintes numériques. Nous obtenons alors une mesure $x/m$, avec $x$ le nombre d'empreintes numériques identiques et $m$ la taille du CFG référent comme ci-dessus.

Ainsi, par cette méthode, nous pouvons caractériser le programme Android inconnu comme étant un maliciel connu du dictionnaire, une variante d'un maliciel du dictionnaire ou un programme a priori sain car sans correspondance dans le dictionnaire.

\section{Sélection des blocs d'opcode caractéristiques}
\label{bdopcodes}

\subsection{Analyse des Ransomware}

\begin{figure}[h]
    \centering
        \includegraphics[scale=0.5]{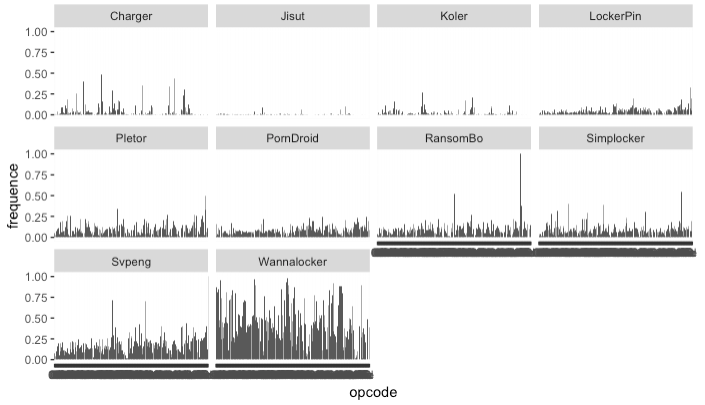} 
        \caption{ \label{treble}Fréquence des ngrams de taille 1 - ransomware}
\end{figure}

L'analyse de la fréquence des opcodes appartenant aux malwares de la catégorie ransomware montre un schéma~\fig{treble} caractéristique à chaque famille. Similaire à l'étude des langues, nous avons considéré les opcodes comme un alphabet et les méthodes comme des phrases. Nous avons donc utilisé les méthodes propres au "Natural Language Processing"(NLP) notamment le score TF/IDF qui permet d'identifier les ngrams importants.

TF représente la nombre de fois qu'un terme apparait dans un document. Il peut s'exprimer comme ceci:

\[TF(word) = \frac{Count(Word)}{\sum_{i=0}^{n} Count(Word_i)}\]

Count(Word) représente le nombre de fois que le mot "word" est présent dans le document.

IDF est la fréquence inverse du document, il s'agit d'une mesure de l'importance d'un terme dans un document.

\[IDF(Word) = log\frac{\left | D \right |}{\left | {d_j: t_{Word} \in d_j} \right |}\]

$|D|$ est le nombre total de document dans le corpus.

$| {d_j: t_{Word} \in d_j} |$ est le nombre de documents où le terme apparaît.

Le score TF/IDF s'obtient en multipliant tf et idf.

Les ngrams sont intéressants comme éléments caractéristiques. Par exemple, le ngram de taille 4 "22 07 07 07" est présent à 97\% au sein de la famille Wannalocker, le ngram de taille 4 "76 e0 a0 10" n'est quant à lui présent qu'au sein de la famille Wannalocker. Le ngram de taille 8 "07 69 07 69 07 69 07 69" n'est présent qu'au sein de la famille Simplocker. Ces observations vont servir de base à la création des séquences qui vont être utilisées lors de l'apprentissage.

Notre approche de comparaison des graphes est tributaire d'une base de données de graphes caractéristiques. Nous nous sommes appuyé sur le jeu de données CICAndMal2017~\cite{dataset} mis à disposition par le "Canadian Institute for Cybersecurity"\footnote{https://www.unb.ca/cic/datasets/invesandmal2019.html}. Il s'agit d'une collection regroupant 10854 programmes dont 4354 malwares. Les malwares sont regroupés par famille appartenant à l'une des catégories suivante:
\begin{itemize}
    \item Adware
    \item Ransomware
    \item Scareware
    \item SMS Malware
\end{itemize}

Chaque catégorie est composée de 10 familles de malwares et chaque famille contient en moyenne 10 programmes.
Les programmes sains sont découpés en années, un jeu de données pour 2015, 2016 et 2017. 
Notre expérimentation va s'effectuer sur les malwares de type Ransomware et sur les programmes sain datant de 2015 (1500 apk).

\subsection{Génération des séquences}

Nous procédons à l'extraction pour chaque malware des méthodes\footnote{les méthodes inhérentes à l'API de Google ne sont pas traitées} sous la forme d'un vecteur composé des opcodes.

\begin{figure}[h]
    \centering
        \includegraphics[scale=0.4]{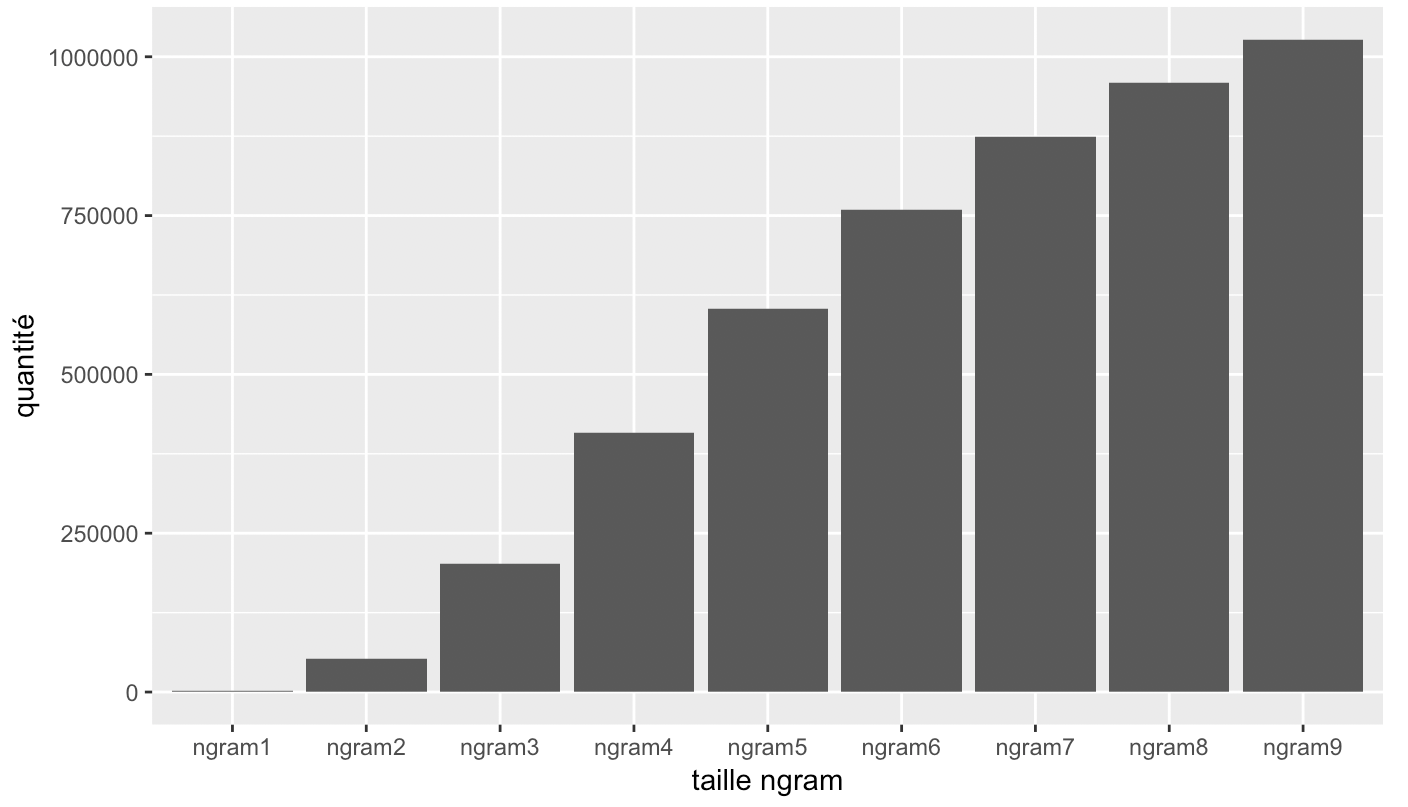} 
        \caption{ \label{ng}Quantité de ngrams en fonction de la taille}
\end{figure}

Pour chaque méthode, nous procédons à la sélection des ngrams caractéristiques en nous appuyant sur le score TF/IDF. Cette sélection dépend également de la taille des ngrams. La~\fig{ng} présente le nombre de ngrams en fonction de leur taille. Le nombre de ngrams uniques et potentiellement intéressants est plus important pour des ngrams de grande taille (supérieur à 4). 

\begin{figure}[!h]
\begin{multicols}{2}
    \includegraphics[width=200px]{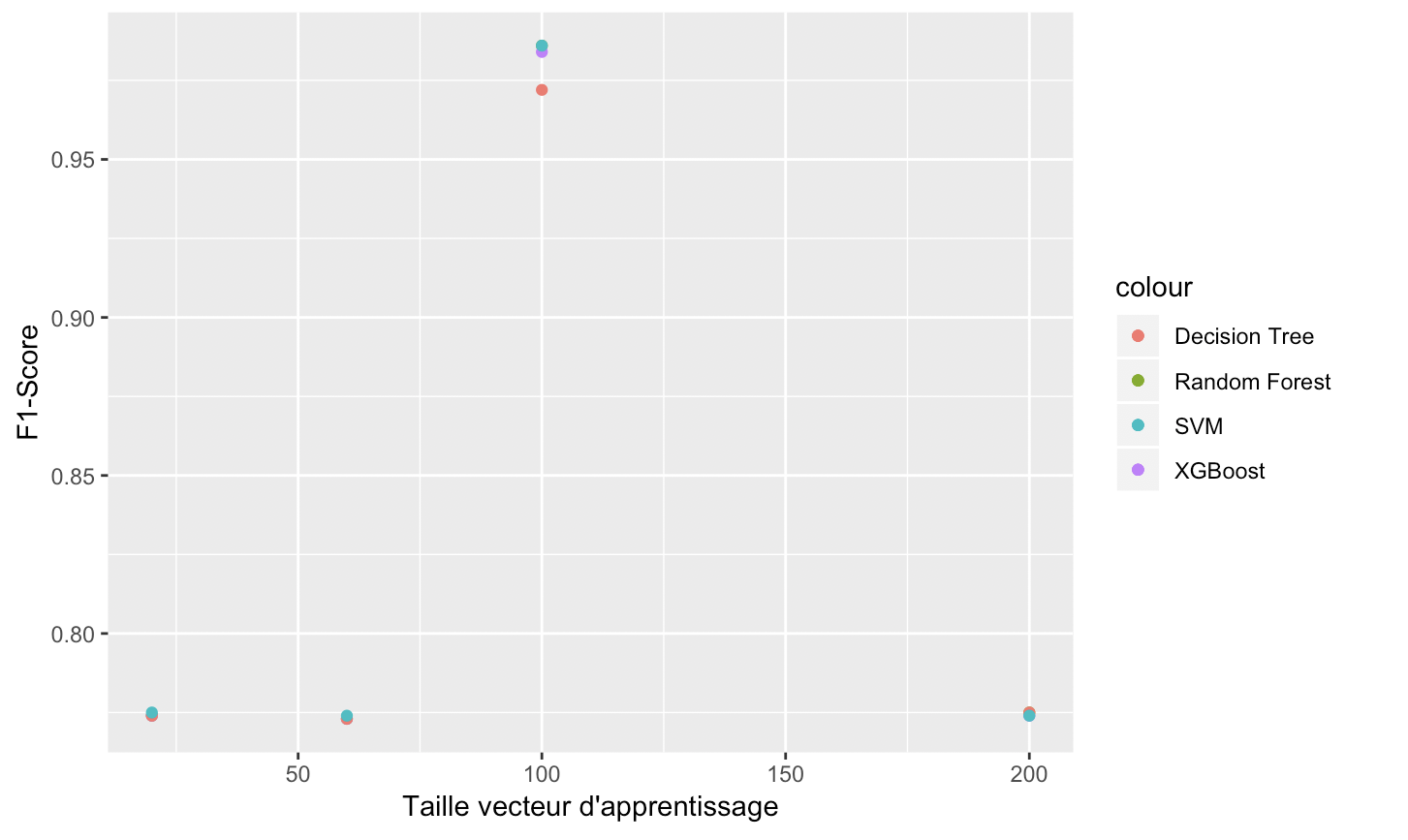}\par 
    \includegraphics[width=200px]{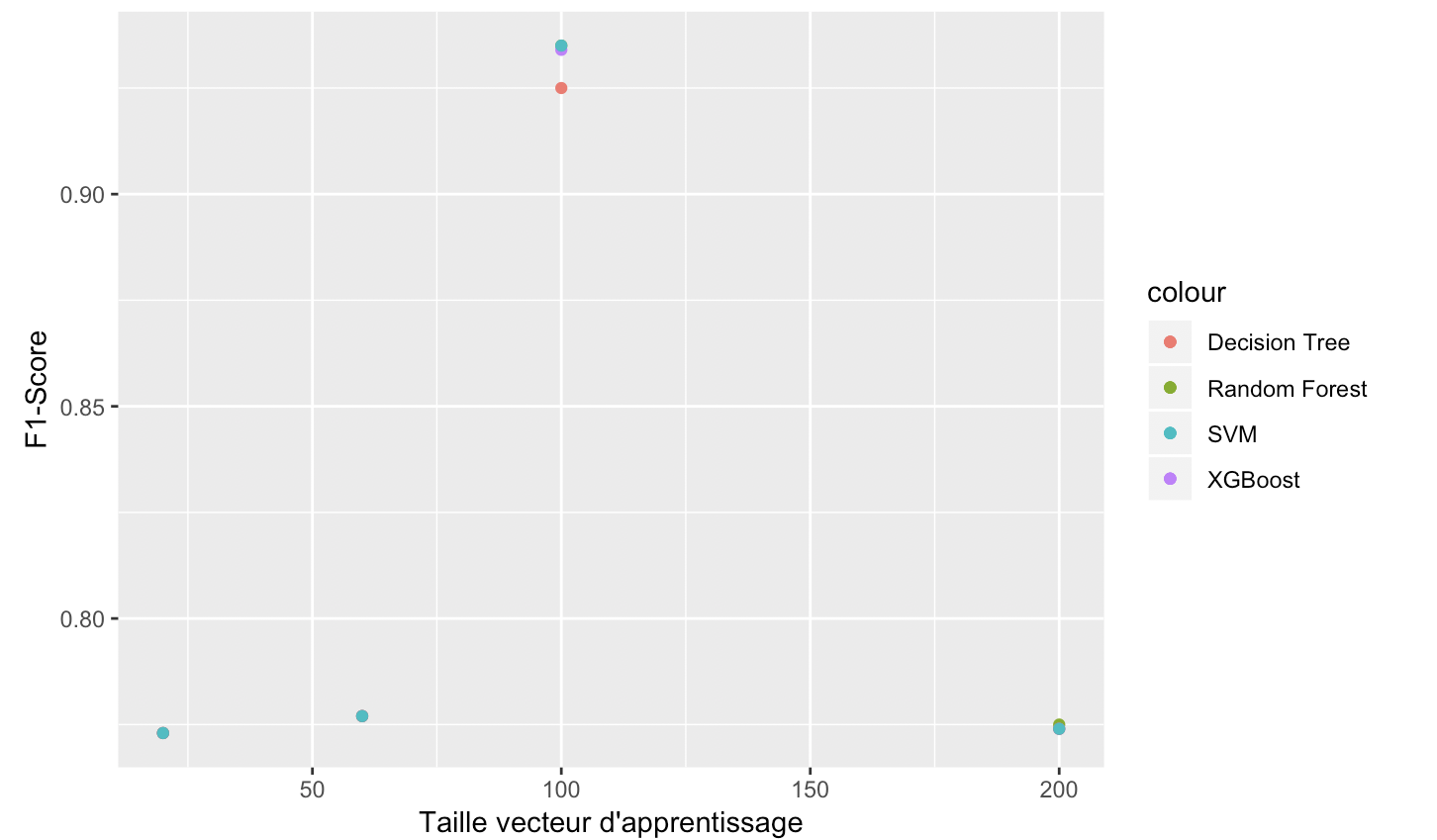}\par 
    \end{multicols}
\begin{multicols}{2}
    \includegraphics[width=200px]{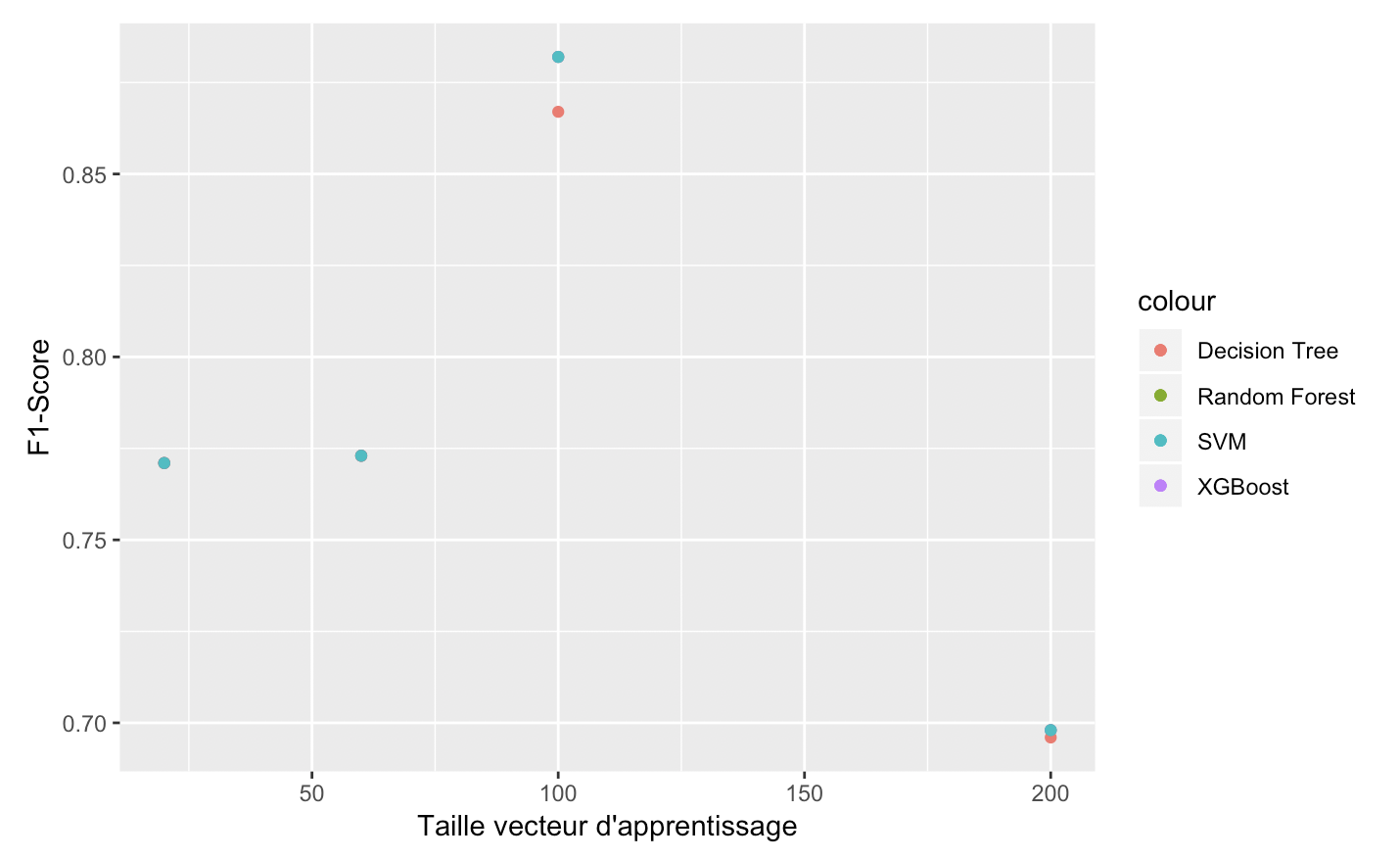}\par
    \includegraphics[width=200px]{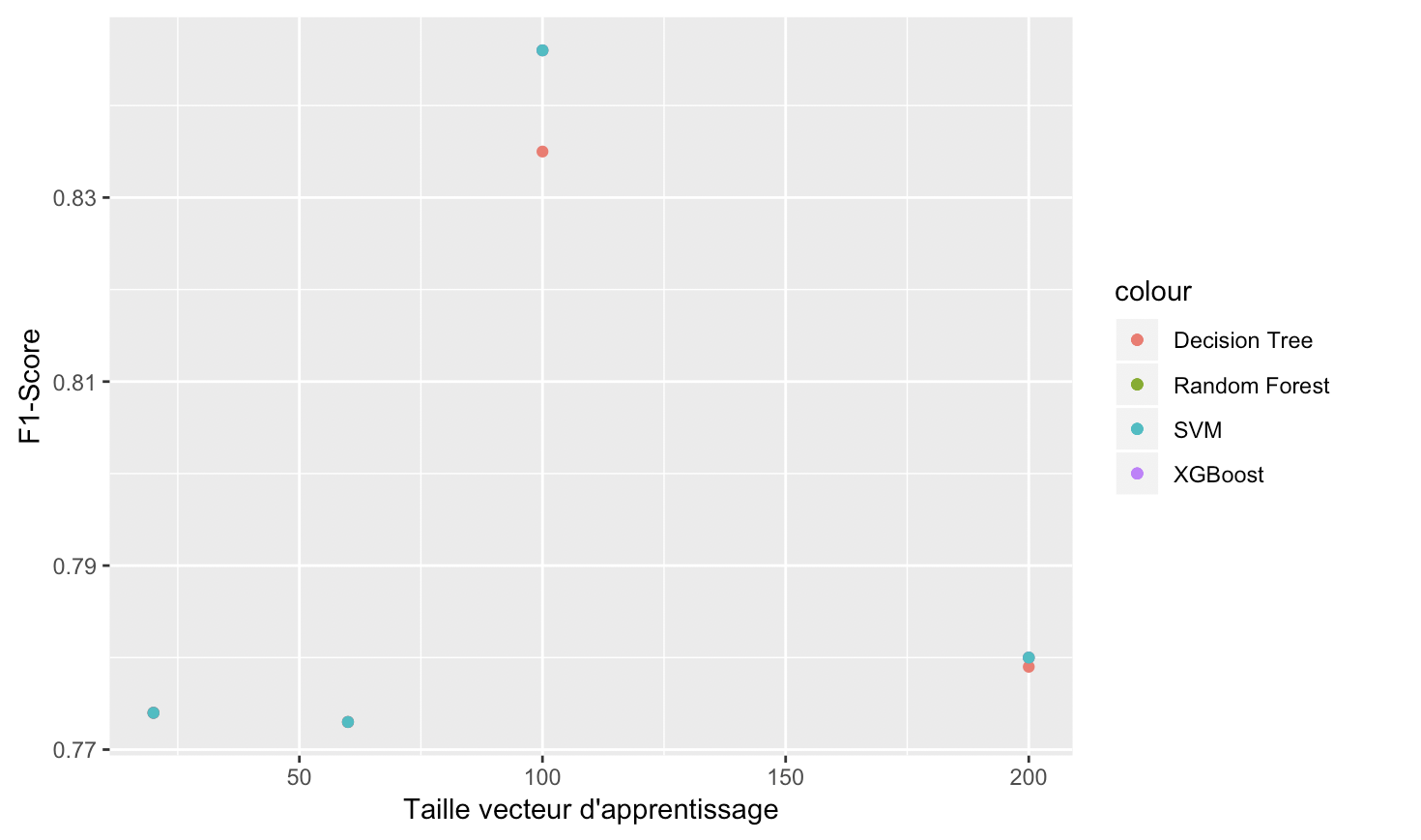}\par
\end{multicols}
\caption{\label{fre}F1-score selon la taille des vecteurs pour les ngrams 2 à 5}
\end{figure}

Nous avons également testé~\fig{fre} pour les ngrams d'une taille comprise entre 2 et 5 différentes tailles de vecteurs.  La taille de 100 représente les meilleurs résultats.

Nous sélectionnons donc les 100 ngrams les plus discriminants afin de créer un vecteur de référence.
Ensuite chaque méthode va être transformée en vecteur indiquant la présence des ngrams selectionnés au sein de chaque méthode. Les programmes sains sont traités de la même manière. Une fois toutes les méthodes transformées, nous ajoutons les classes (0: sain, 1: malware) afin de préciser la provenance et d'utiliser des algorithmes d'apprentissage supervisé. 

\subsection{Apprentissage}

Notre choix s'est porté sur les algorithmes supervisés suivants (Random Forest, XGBoost, Decision Tree, SVM et KNN). Ces derniers sont appropriés à la prédiction d'une classe à partir d'un ensemble de données. Les calculs ont été automatisés via l'outil Dataiku.

Nous avons testé les algorithmes suivants pour des ngrams d'une taille de 1 à 9 et avons obtenus le F1-score suivant. Le jeu de données utilise 80 \% du jeu pour l'apprentissage et 20\% pour les tests. Une validation croisée de 10 est utilisée pour tous les tests.

\begin{table}[h]
 \hspace*{-0.5in}
\begin{tabular}{llllllllll}
                                          &                                     &                                     & \multicolumn{3}{l}{Taille des ngrams}                                                                           &                                     &                                     &                                     &                                     \\ \cline{2-10} 
\multicolumn{1}{l|}{Algorithmes}          & \multicolumn{1}{l|}{1}              & \multicolumn{1}{l|}{2}              & \multicolumn{1}{l|}{3}              & \multicolumn{1}{l|}{4}              & \multicolumn{1}{l|}{5}              & \multicolumn{1}{l|}{6}              & \multicolumn{1}{l|}{7}              & \multicolumn{1}{l|}{8}              & \multicolumn{1}{l|}{9}              \\ \hline
\multicolumn{1}{|l|}{Random Forest}       & \multicolumn{1}{l|}{\textbf{0.923}} & \multicolumn{1}{l|}{\textbf{0.986}} & \multicolumn{1}{l|}{\textbf{0.935}} & \multicolumn{1}{l|}{\textbf{0.882}} & \multicolumn{1}{l|}{\textbf{0.846}} & \multicolumn{1}{l|}{\textbf{0.829}} & \multicolumn{1}{l|}{\textbf{0.820}} & \multicolumn{1}{l|}{\textbf{0.794}} & \multicolumn{1}{l|}{\textbf{0.796}} \\ \hline
\multicolumn{1}{|l|}{XGBoost}             & \multicolumn{1}{l|}{0.919}          & \multicolumn{1}{l|}{0.984}          & \multicolumn{1}{l|}{0.934}          & \multicolumn{1}{l|}{0.882}          & \multicolumn{1}{l|}{\textbf{0.846}} & \multicolumn{1}{l|}{\textbf{0.829}} & \multicolumn{1}{l|}{\textbf{0.820}} & \multicolumn{1}{l|}{0.793}          & \multicolumn{1}{l|}{\textbf{0.796}} \\ \hline
\multicolumn{1}{|l|}{Decision Tree}       & \multicolumn{1}{l|}{0.916}          & \multicolumn{1}{l|}{0.972}          & \multicolumn{1}{l|}{0.925}          & \multicolumn{1}{l|}{0.867}          & \multicolumn{1}{l|}{0.835}          & \multicolumn{1}{l|}{0.821}          & \multicolumn{1}{l|}{0.813}          & \multicolumn{1}{l|}{0.787}          & \multicolumn{1}{l|}{0.793}          \\ \hline
\multicolumn{1}{|l|}{KNN} & \multicolumn{1}{l|}{0.916}          & \multicolumn{1}{l|}{0.982}          & \multicolumn{1}{l|}{0.934}          & \multicolumn{1}{l|}{0.882}          & \multicolumn{1}{l|}{\textless 0.5}  & \multicolumn{1}{l|}{\textless 0.5}  & \multicolumn{1}{l|}{\textless 0.5}  & \multicolumn{1}{l|}{\textless 0.5}  & \multicolumn{1}{l|}{\textless 0.5}  \\ \hline
\multicolumn{1}{|l|}{SVM}                 & \multicolumn{1}{l|}{0.917}          & \multicolumn{1}{l|}{\textbf{0.986}} & \multicolumn{1}{l|}{\textbf{0.935}} & \multicolumn{1}{l|}{0.882}          & \multicolumn{1}{l|}{\textbf{0.846}} & \multicolumn{1}{l|}{\textbf{0.829}} & \multicolumn{1}{l|}{\textbf{0.820}} & \multicolumn{1}{l|}{0.793}          & \multicolumn{1}{l|}{\textbf{0.796}} \\ \hline
\end{tabular}
 \caption{F1 score pour les algorithmes en fonction de la taille des ngrams pour taille de vecteur de 100} \label{tabl}
\end{table}

Random forest et SVM présentent le meilleur score. A l'inverse KNN est devenu rapidement inefficace dès une taille de ngram supérieure ou égale à 5.

Avec cette approche, nous pouvons déterminer les méthodes intéressantes parmi plusieurs centaines d'apk afin de constituer notre base de données pour la recherche d'isomorphisme.

\section{Expérimentation}

L'expérimentation de notre méthode a été réalisée en considérant deux configurations complémentaires. Dans un premier temps, nous avons cherché à tester la performance et la robustesse de notre méthode. Pour cela, il nous faut pouvoir mesurer la performance et le rappel sur un jeu de donnée suffisamment important. Considérant la difficulté à trouver un grand nombre de variantes d'un même maliciel, nous avons décidé de créer nous même un jeu de données ad hoc, sans maliciels, mais avec un code considéré pour notre cause comme infecté. Nous avons introduit trois variantes de ce code dans dix applications Android. Cela nous a permis de créer l'équivalent de 30 applications "infectées", toutes des variantes du même "maliciel". Nous avons complété ce jeu avec les 10 applications non modifiées et 100 autres applications. Cette expérimentation de laboratoire nous a permis de valider le concept de notre approche.

Dans un second temps, nous avons cherché à tester notre méthode sur un jeu de données réelles. Pour cela, nous avons récupéré des variantes des maliciels DroidJack et Opfake, 10 au total. Nous avons aussi constitué une base de données de 100 programme sains et autant de maliciels. Cette expérimentation a permis de tester notre méthode dans des conditions réelles.

\subsection{Jeu de données de laboratoire}

Afin de tester la validité de notre approche, nous avons constitué un jeu de données de laboratoire simulant le comportement de maliciels, sans en être. Pour cela, nous avons codé une classe singleton permettant d'envoyer automatiquement et sans consentement un SMS a un numéro de téléphone. Deux variantes de cette classe ont été implémentées qui modifient légèrement le code précédent par ajout ou suppression d'une ligne.

Nous avons sélectionné 10 applications Android FOSS dont le code source est ouvert. Nous avons créé un jeu de 30 applications considérées comme "infestées" en incluant indépendamment chacune des trois variantes à ces applications.

Nous avons aussi créé une catégorie d'application considérées comme "saines", constituée des 10 applications Android FOSS non modifiées et de 100 apk testés comme étant saines sur Virustotal~\cite{virustotal}. 

L'intérêt de ce jeu de données de laboratoire est de pouvoir décortiquer les résultats variante par variante. Nous avons donc constitué trois dictionnaires, chacun basé sur une unique variante. Les résultats obtenus par notre méthode pour chacun de ces trois dictionnaires sont présentés~\ta{tab-labo}. Ces résultats ont été obtenus en choisissant comme seuil de correspondance qu'un isomorphisme de sous-graphe ait bien été trouvé dans le programme candidat et qu'au moins la moitié des empreintes numériques correspondent. Ces résultats valident pleinement notre méthode, mais le seuil de correspondance ne doit bien évidement pas être choisi trop haut.


\begin{table}
  \centering
  \begin{tabular}{|l|ccc|}
    \hline
    & Précision & Rappel & F-mesure \\
    \hline
    Variante 1 & 1 & 1 & 1 \\
    Variante 2 & 1 & 1 & 1 \\
    Variante 3 & 1 & 1 & 1 \\
    \hline
  \end{tabular}
  \caption{Résultats obtenus sur le jeu de données de laboratoire constitué de 30 applications considérées comme infestées et 100 applications ne l'étant pas}
  \label{tab-labo}
\end{table}

\subsection{Jeu de données réelles}

Afin de valider notre méthode dans des conditions réelles, nous avons constitué un jeu de données d'application réellement saines et de maliciels. Pour cela, nous avons récoltés 100 exemplaires infestés provenant de la plateforme collaborative Koodous~\cite{koodous}. Parmi ces exemplaires, nous en avons sélectionné deux en particulier et leurs variantes : DroidJack et Opfake. Nous avons récolté 6 variantes de DroidJack et 4 variantes de Opfake. Avec les 100 logiciels sains dont nous avons parlé à la section précédente, nous avons donc constitué un jeu de données réelles de 100 applications saines et 100 maliciels dont 10 variantes issues de deux maliciels.

En ce basant sur ce jeu de données, nous avons réalisés trois expérimentations :
\begin{itemize}
\item Création d'un dictionnaire basé uniquement sur une variante de DroidJack ;
\item Création d'un dictionnaire basé uniquement sur une variante de Opfake ;
\item Création d'un dictionnaire à partir des 100 maliciels connus.
\end{itemize}

Les résultats de ces trois expérimentations sont présentées~\ta{tab-reel}. L'expérimentation basés sur les 6 variantes connues de DroidJack montre que nous avons obtenus deux faux positifs. Les correspondances entre les CFG de ces deux faux positifs et ceux de DroidJack étant complètent, il n'est pas impossible que ces deux faux positifs aient été mal classés dans Koodous. 

\begin{table}
  \centering
  \begin{tabular}{|l|ccc|}
    \hline
    & Précision & Rappel & F-mesure \\
    \hline
    DroidJack & 0,98 & 1 & 0,99 \\
    Opfake & 1 & 1 & 1 \\ 
    Ensemble des maliciels & 1 & 1 & 1 \\
    \hline
  \end{tabular}
  \caption{Résultats obtenus sur le jeu de données réelles constitué de 100 applications saines et 100 maliciels}
  \label{tab-reel}
\end{table}

\section{Conclusion}

Android est le système d’exploitation le plus répandue au monde. Disponible sur
de nombreux terminaux, il fait partie intégrante de notre quotidien personnel
comme professionnel. De fait, il constitue une cible idéale pour les attaquants.
La détection fiable des logiciels malveillants, zero-day ou connus, ainsi que leurs
variantes, est un enjeux majeur pour la sécurité.

Nous proposons pour répondre à cet enjeux une approche innovante, basée sur
une analyse statique du logiciel afin d’éviter les difficultés de mise en oeuvre
des analyses dynamiques ainsi que leurs nombreuses parades. Pour détecter un
maliciel, notre méthode va réaliser une comparaison non seulement sémantique,
basée sur les opcodes, mais aussi structurée, basée sur les liens logiques. Elle est
complétée par un dictionnaire de blocs de codes infectés construit par appren-
tissage automatique.

Plus précisément, notre méthode de détection de maliciels Android est basée sur
la construction de graphes de flots de contrôle (CFG) à partir des opcodes de
la machine virtuelle Dalvik. La caractérisation d’un nouveau maliciel se fait en
comparant ses CFG d’opcodes via la recherche d’isomorphisme de sous-graphes
avec une base de donnée de CFG caractéristiques soigneusement sélectionnés
parmi une base de maliciels connus. Ces derniers CFG sont sélectionnés par un
processus d’apprentissage automatique parmi l’ensemble des CFG extraits de la
base de maliciels.

Les résultats réalisés à partir de deux jeux de données, un jeu de données de laboratoire construit de manière ad hoc et un jeu de données réelles, valident notre approche en obtenant d'excellents scores de F-mesure. Le score TF/IDF s'est montré intéressant dans la recherche de méthodes malignes.

\bibliographystyle{plainurl}
\bibliography{Detection_maliciel_Android}

\end{document}